\documentclass{mem}
\usepackage{natbib}\usepackage{txfonts}\usepackage{balance}
\usepackage{graphicx}
\usepackage[a4paper,breaklinks,dvipdfm]{hyperref}
\idline{75}{1}
\begin{document}
\def\teff{$T\rm_{eff }$}
\def\kms{$\mathrm {km s}^{-1}$}

\title{From Helio- to Asteroseismology and the progress in stellar physics
}


\author{
M. P. \,Di Mauro\inst{1} 
          }

  \offprints{M. P. Di Mauro}

\institute{$^1$
INAF--IAPS,
Istituto di Astrofisica e Planetologia Spaziali, Via del Fosso del Cavaliere 100,
I-00133 Roma, Italy
\email{maria.dimauro@inaf.it}
}

\authorrunning{Di Mauro}

\titlerunning{From Helio- to Asteroseismology}

\abstract{
During the last decades, numerous observational and theoretical efforts in
 the study of solar oscillations, have brought to a detailed knowledge of
the interior of the Sun.
While this discipline has not yet exhausted its resources and scientists are
 still working on further refinements of the solar models and to solve the numerous still open questions, Asteroseismology, which aims to infer the structural properties of stars which display
multi-mode pulsations, has just entered in its golden age.
In fact, the space missions
 CoRoT and {\it Kepler} dedicated to the observation of stellar oscillations, have
 already unveiled primary results on the structural properties of the stars producing a revolution in the way we study the stellar
interiors.

Here, the modern era of Helio- and Asteroseismology
is reviewed with emphasis on results obtained for the Sun and its
solar-like counterparts.
}
\maketitle{}

\section{Introduction}

\begin{figure} 
\centering 
\includegraphics[width=0.9\linewidth]{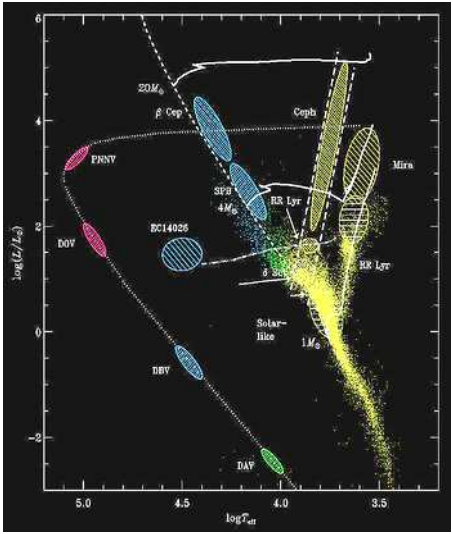} 
\caption{\footnotesize HR diagram showing several classes of pulsating stars. Among them we can identify the classes of solar-like stars, $\delta$-Scuti stars just on or little above the ZAMS, classical pulsators and white-dwarf stars along the cooling sequence (from Christensen-Dalsgaard and Dziembowski 2000)} 
\label{HRall} 
\end{figure}

Helio- and Asteroseismology, whose etymologies have ancient Greek roots, indicate the study of the internal structure and dynamics of the Sun and
other stars from observations of their resonant vibrations. These vibrations – or oscillations – 
manifest themselves in small motions of the visible surface of the star and, like the seismic waves generated by earthquakes in the Earth,  
  provide us with valuable information about the pervaded internal layers. 
Oscillations have several advantages over 
all the other observables:
pulsational instability has been detected in 
stars in all the
evolutionary stages and of different spectral type from main-sequence to the
white dwarf cooling sequence (see HR diagram in Fig.~\ref{HRall});
frequencies of oscillations can be measured with high accuracy and
 depend in very simply way on the equilibrium structure of the model; 
different modes propagate through different layers of the interior of a star.
Thus, a sufficiently rich spectrum of observed resonant modes
 permit the probing of internal 
conditions and lead to the testing and revision of our theories of stellar
structure and evolution. Since a correct understanding of stars' evolution is a corner-stone of
modern astrophysics, these endeavors are fundamentally important to the whole of astrophysics.

Stellar pulsations may be distinguished in self-excited oscillations and stochastic oscillations, according to their own driving mechanism.
The self-excited oscillations, observed in classical pulsators,
arise from a perturbation to the energy flux resulting in 
a heat-engine mechanism converting thermal into mechanical energy. 
If the perturbations are associated with opacity rapid variations, the driving mechanism is known as $\kappa$-mechanism.
Stochastic oscillations, so-called solar-like oscillations, are excited by turbulent convection like in the Sun, and
are predicted for all
 main-sequence and 
postmain-sequence stars cool enough to harbor an outer convective envelope.
 The present paper provides 
a general overview on the recent observational and
theoretical successes obtained on the Sun and
other solar-type stars thanks to
the possibility of handling large sets of accurate oscillation
 frequencies. The author will comment on analogies and differences of using
and adapting to other stars techniques and 
 methods developed for
the Sun.
 
\section{Properties of solar-type pulsators} 

The pulsations supported in a star can be classified in pressure or acoustic 
waves and gravity waves, which form the classes of 
p and g modes respectively, named after the force that acts to restore the stellar equilibrium.

\begin{figure} 
\includegraphics[width=1.0\linewidth]{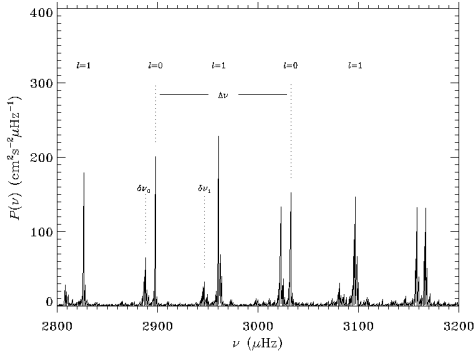} 
\caption{\footnotesize The oscillation power spectrum of the Sun obtained by BiSON \citep[from][]{CD05}.} 
\label{spec} 
\end{figure}

Solar-like oscillations are generally p modes of high radial order which
can be well described in terms of the asymptotic theory (Tassoul 1980). In the asymptotic approximation 
the oscillation frequencies $\nu_{n,l}$ of acoustic modes,   
 characterized by radial order $n$ and harmonic degree $l$   
should satisfy the relation:   
\begin{equation}   
\nu_{n,l}=\Delta\nu\left(n+\frac{l}{2}+\alpha+\frac{1}{4} \right)   
+\epsilon_{n,l} \; ,   
\label{eq1}   
\end{equation}   
where $\alpha$ is a function of the frequency determined by the
properties of the
surface layers, $\epsilon_{n,l}$   
 is a small correction which depends on the conditions in the stellar core.
$\Delta\nu$ is the inverse of    
the sound travel time across the stellar diameter: 
\begin{equation}
\Delta\nu={\left(2\int_{0}^{R}\!\frac{{\rm d}r}{c}\right)}^{-1}, 
\end{equation}
where $c$ is the local speed of sound at radius $r$ and $R$ is the
photospheric stellar radius. To first approximation,
Eq. \ref{eq1} predicts that acoustic spectra should show
a series of equally spaced peaks between p modes of same degree,
whose frequency separation is
the so called large separation which is
approximately equivalent to
$\Delta\nu$:
\begin{equation}
\Delta\nu \simeq\nu_{n+1,l}-\nu_{n,l}\equiv\Delta\nu_{l}\,.
\label{EQ_3}
\end{equation}
In addition, the spectra are characterized by another series of peaks, whose narrow separation is $\delta\nu_{l}$, known as small separation:
\begin{equation}
\delta\nu_{l}\equiv \nu_{n,l}-\nu_{n-1,l+2}=(4l+6){\rm D}_{0}
\label{EQ_4}
\end{equation}
where
\begin{equation}
{\rm D}_{0}=-\frac{\Delta\nu}{4\pi^{2}\nu_{n,l}}\left[\frac{c(R)}{R}-
\int_{0}^{R}
\frac{{\rm d}c}{{\rm d}r}\frac{{\rm d}r}{r}\right] \, .
\end{equation}
As an example, the oscillation spectrum of the Sun is plotted in Fig. \ref{spec}.
$\Delta \nu$, and hence the general spectrum of acoustic modes,   
scales approximately as the square root of the mean density, that is,   
for fixed mass, as $R^{-3/2}$: as the star evolves the large separation
decreases with the increase of the radius. 
 On the other hand, the small frequency separation is sensitive to the sound-speed gradient in the core, which in turn is sensitive to the chemical composition gradient in   
 central regions of the star and hence to    
 its evolutionary state. 

\begin{figure}
\includegraphics[width=1.0\linewidth]{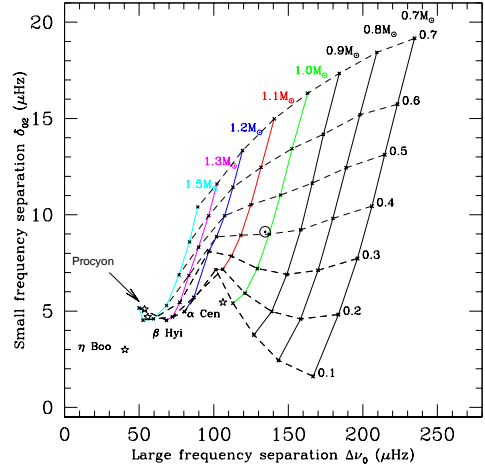} 
\caption{\footnotesize C-D digram showing large separation versus small separation for a series of evolutive
  models calculated for different masses, but fixed metallicity. Solid lines are evolutive tracks
  for decreasing core's hydrogen abundance. Dashed black lines indicate models with same core's hydrogen content decreasing from $X_c=0.7$(ZAMS) at the top, to $X_c=0.1$ at the bottom.  The
Sun is marked by $\odot$. Black stars locate ground based observations of known solar-type stars. Courtesy of D. Cardini.} 
\label{CD} 
\end{figure}

The determination of the large and small frequency separations
can provide asteroseismic inferences on the mass and the age of
main-sequence and post-main-sequence solar-type stars, using the
so-called seismic diagnostic 'C-D diagram' (Christensen-Dalsgaard
1988) like Fig. \ref{CD}.

Asymptotic properties have been derived also for low-degree g-modes with high radial order.
Tassoul's theory (Tassoul 1980) shows that in the asymptotic regime the g-modes of same harmonic degree are nearly uniformly spaced in period:
 \begin{equation}
\Delta P_{n,l}=\frac{N_0}{\sqrt{l(l+1)}}(n+\alpha_{l,g}),
\label{EQ_10}
\end{equation}
where
\begin{equation}
N_0=2\pi^2\left(\int_{r_1}^{r_2} N\frac{dr}{r}\right)^{-1}
\end{equation}
$N$ is the buoyancy frequency, $r_1$ and $r_2$ define the region of propagation of the g modes and $\alpha_{l,g}$ is the phase term which depends on the details of the boundaries of the trapping region.

\begin{figure}
\includegraphics[width=.7\linewidth, angle=270]{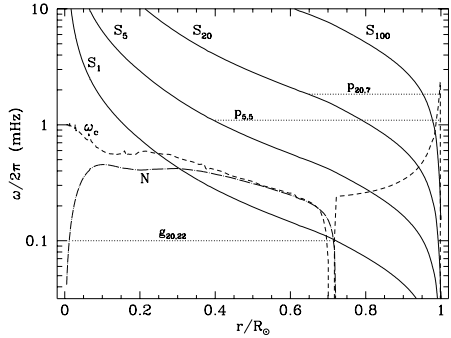}
\caption{\footnotesize Propagation diagram obtained for a solar standard model. The horizontal lines indicate the trapping regions  for a g mode with $l=20$ and $n=22$, and two p modes 
with ($l=5 \, ,n=5$) and ($l=20\,,n=7$)}
\label{prop}
\end{figure}

The regions of propagation 
of p and g modes can be well illustrated by a propagation diagram, like in Fig. \ref{prop}. The trapping region of g modes is delimited by the buoyancy frequency $N$, while
the Lamb frequency $S_l$ and
the acoustical cut-off frequency $\omega_c$  define the region of propagation of the p modes. 
In a main sequence star, like the Sun,
the gravity modes are trapped at low frequencies in the radiative
interior, while the acoustic modes propagate
at high frequencies through the convective zone up to the surface. 
Outside these regions
 the waves are evanescent and do not show oscillatory character in space and their amplitude decays exponentially.

\begin{figure}
\centering
\includegraphics[width=1.0\linewidth]{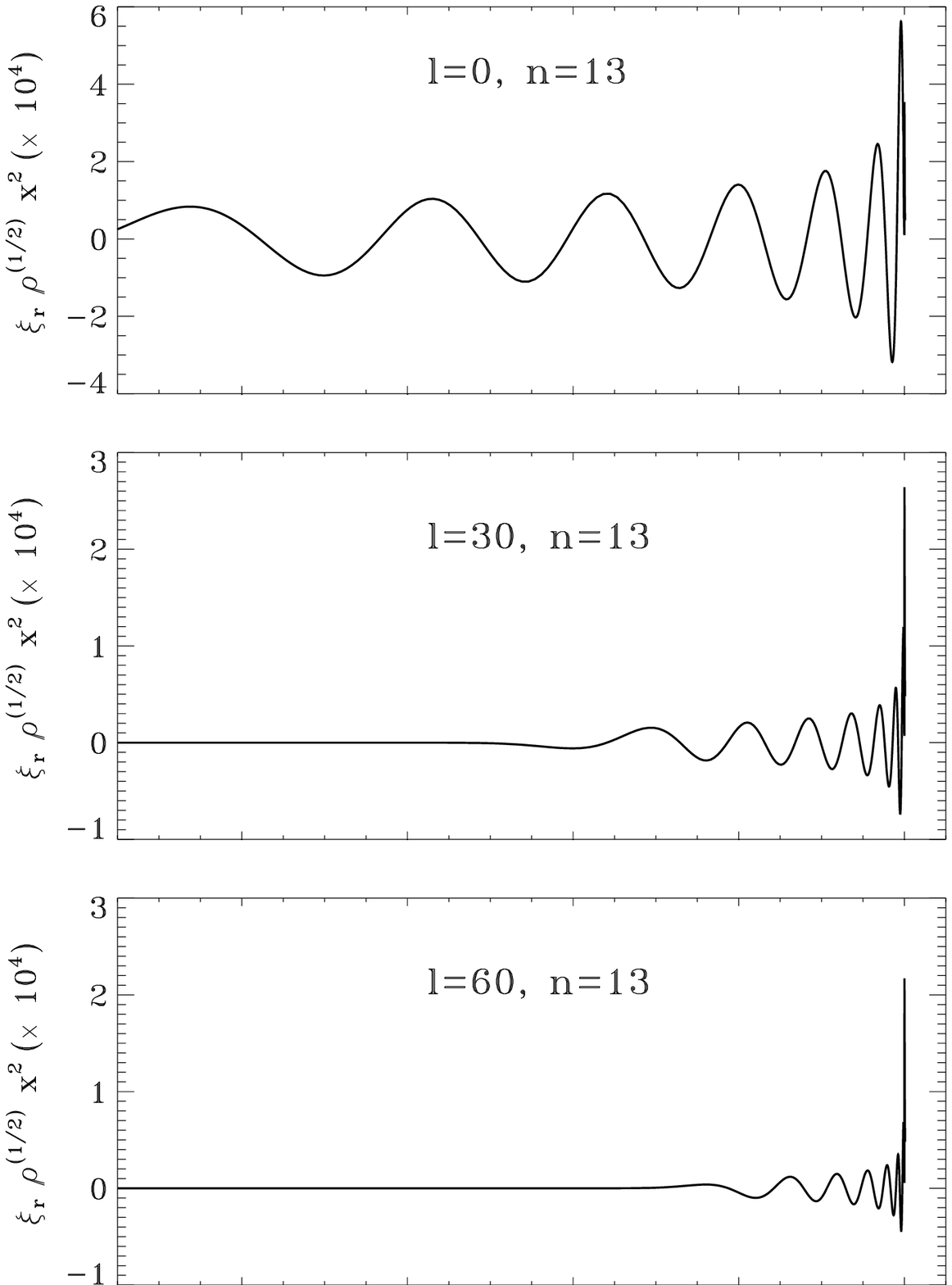}
\vspace{0.5cm}
\caption{\footnotesize Eigenfunction for p modes with different harmonic degree
as function of the fractional radius $x=r/R_{\odot}$ 
in a standard solar model. Here, the oscillation behavior is enhanced, by 
scaling the eigenfunctions with the square root of the density and the squared fractional radius.}
\label{aut}
\end{figure}

Figure \ref{aut} shows oscillation eigenfunctions for a selection of p modes with different harmonic
degree calculated for a standard solar model.
The lower is the harmonic degree $l$,
the deeper is located the turning point of the acoustic mode.
Radial acoustic modes with $l=0$ penetrate to the centre, while the modes of
 highest harmonic degree are trapped in 
the outer layers.

\section{Seismology of the Sun}

During the last decades, Helioseismology has dramatically changed our understanding of the structure of the Sun, its internal dynamics and the temporal evolution of the solar interior. 
This progress has been possible due to the development of the study of the solar oscillations and the large quantity of observed pulsation modes detected by several
helioseismic experiments. The IRIS \citep{fossat}, the GONG (Global Oscillations Network Group) \citep{ha96} and the BiSON \citep{chaplin96} networks,
consisting of a number of observing stations worldwide located at different latitudes, has allowed to monitor our star without temporal interruption.
 But the great success of Helioseismology arrived with the launch of the ESA/NASA SOHO spacecraft in 1996 and its three instruments, the Solar Oscillations Imager / Michelson Doppler Imager (SOI/MDI) \citep{sc95} , the Global Oscillations at Low Frequency (GOLF) \citep{gabriel5}  and the Variability of solar Irradiance and Gravity Oscillations (VIRGO) \citep{frolich}.

Solar oscillations can be studied through two different techniques: global Helioseismology and local Helioseismology.
"Global" Helioseismology, based on the analysis of mode frequencies reveals large-scale properties of 
the solar structure and dynamics of the Sun;
``local'' Helioseismology, based 
on the use of the travel times of the acoustic waves through the interior
 between different points on the solar surface,
provides three-dimensional 
maps of the sound speed and of the flows in the upper convection zone, to probe local 
inhomogeneities in the sub-surface and surface layers. 

The measurement of thousand of individual oscillations frequencies has allowed to establish the correctness of the standard solar model, to improve the description of the relevant physics like the equation of state, the opacity table,
and the nuclear reactions and to refine some internal details by inclusion of additional effects such as the heavy-elements diffusion, rotation, overshooting etc..
Helioseismology, so far, has shown that 
the solar structure is remarkably close to the predictions of the standard solar model and that 
the interior can be probed with sufficiently high spatial resolution to 
allow investigation on the internal dynamics, the element abundances in the solar envelope (e.g., 
\citealt{ko92,dz92,ba97}) and also the presence of dark matter in the core (e.g., \citealt{turck}).

On the other hand, Helioseismology pointed out
that the structure of the convection zone and of the 
near-surface region of the solar models remains still quite uncertain.
In fact, the
solar models are based
on several, perhaps questionable, assumptions about the physical 
properties of matter in stars and
 the computation of models involves a number of simplifying hypotheses such as 
the treatment of convection, generally approximated by mixing-length theory, while
 the dynamical effects of the turbulent pressure are often neglected.

Finally, Helioseismology contributed to solve the well known 'neutrino-problem': the
agreement of the solar models with helioseismic data strongly suggested that the solution of the neutrino problem had to be sought not in the physics of the solar interior but rather in the physics of the solar neutrino.

At present, two new 
helioseismological experiments are
devoted to the measurements from space: 
SDO, Solar Dynamics Observatory \citep{sc12}, a NASA mission to understand the temporal variation of the solar magnetic field, the dynamical processes and their impact on Earth (launched in 2008);
PICARD \citep{picard}, a CNES mission to study the Earth climate and Sun variability relationship (launched in 2009). 
For the future there is a high expectation for
SO, Solar Orbiter \citep{SO}, an ESA-NASA cooperation satellite 
to study the polar regions and the side of the Sun not visible from Earth. 

\subsection{The solar dynamics}

It is well known, and easily observed at the photosphere, that the Sun is 
slowly rotating.
 The rotation breaks the spherical symmetry and splits the frequency of each oscillation mode of harmonic degree $l$ into $2l+1$ components, known as ``splittings''. By applying standard perturbation theory to eigenfrequencies, it can be shown that the rotational splitting for each mode is directly related to the rotation rate $\Omega(r, \theta)$ inside a star,
where $r$ is the radius and
 $\theta$ is the co-latitude.
 The dependence of the splittings on angular velocity can be used in a inverse problem to probe the Sun's internal differential rotation.

The results about the internal angular velocity of the Sun have
shown that the latitudinal differential rotation observed at the surface persists throughout the convection zone, while the radiative interior rotates almost rigidly at a rate of about $430\,{\rm nHz}$ \citep[e.g.,][]{sc98,dim98}. 
The global dynamo action, responsible for the generation of the solar $22\,{\rm yr}$
magnetic cycle, is induced by the strong toroidal magnetic fields generated by rotation shear in the tachocline,
the thin transition layer from latitude-dependent rotation to nearly independent rotation \citep[e.g.,][]{spi, sc98}.

\subsection{The solar core}
The problem of inferring the core physics remains
one of the most important open questions in Helioseismology.
Low-degree p modes are able to penetrate towards the centre, sampling the core for a relative short time because of the large sound speed there.
Thus, p modes, as opposed to
g modes, are not very sensitive to the features of the core of the Sun.
In addition, 
low-degree data sets obtained by different instruments are not in mutual agreement and give conflicting inversion results in the core \citep{dim98}.
Recently, after the analysis of
10 years of data collected by GOLF, the detection of a g mode was finally announced
by \citet{ga07}. According to the authors,
 only models characterized by
 a core rotation faster than in the rest of the radiative zone are 
 able to reproduce the frequency of such a g mode.
Inversion of set of data, including p-mode splittings observed with MDI and 5 g-mode splittings detected with GOLF, have been performed by \citet{ga11}, leading to a result consistent with a core's rotation from 5 to 7 times
faster than the surface.
These results still wait to be confirmed and verified by use of different approaches of analyses.

\section{Seismology of solar-type stars}

Inferences
of the interior of stars other than the Sun 
appear to be much more complicated and less outstanding in terms of 
achievable results.
The large stellar distances, the point-source character of the stars,  restrict the asteroseismic studies
to the use of small sets of data often characterized by modes with only low harmonic 
degrees ($l\leq 4$).
Nevertheless, over the past decade,
Asteroseismology has developed as a consequence of the very important successes obtained by Helioseismology and thanks to
the improved quality of the seismic observations, from ground-based spectroscopy to the space missions launched in recent years and dedicated to the measure of stellar pulsations outside the atmosphere.

The first dedicated Asteroseismology mission to be launched successfully was MOST (Microvariability and Oscillations of Stars) \citep{walker03}, which has
achieved great success with observations of classical (heat-engine) pulsators. In the realm of solar-like oscillations, controversy was generated
when MOST failed to find evidence for oscillations in Procyon A
(Matthews et al. 2004). 
However, the very recent results by
Huber et al. (2011) on Procyon~A, based on a simultaneous ground-based spectroscopic campaign (Arentoft et al. 2008) and high-precision photometry by the MOST
satellite (Guenther et al. 2008), have revealed
 that the problems rely in the modelling of the convective transport and the wrong estimation of the
oscillation amplitudes and mode lifetimes in stars somewhat more evolved than the Sun (Houdek et al. 1999).

The French-led CoRoT mission, launched in 2006, contributed substantially to observe main-sequence, subgiant and red-giant stars with solar-like oscillations
(e.g., Appourchaux et al., 2008).

But the real revolution in the stellar physics field was produced by
{\it Kepler} which, launched in 2009 with the primary goal to search for
extra-solar planets (Borucki et al. 2010), has
 released photometry time-series data for about $\simeq 150,000$ stars 
enabling to study internal structure and properties of several thousands of pulsating stars \citep[e.g.,][]{chaplin10} including some exoplanet hosts \citep[e.g.,][]{CD10}.

Preliminary asteroseismic studies, such as
the extraction of a rough estimate of the global parameters of a star, are
 possible by using the average properties of the oscillation spectra, such as the large and small separations and the frequency of the maximum oscillation power, $\nu_{max}$. 
The C-D diagram, as introduced in Sec. 2, can be used to determine mass and age of
the stars. In main-sequence stars, the tracks for different masses and ages
are well separated.
For more evolved stars, the tracks converge and the small separation becomes
much less sensitive as a diagnostic. 

Another powerful seismic tool is represented by the use of the scaling laws
provided by \citet{kjeldsen95} and \citet{bedding03}, which
allow to derive stellar mass and radius of stars by knowing the large
separation and the frequency of the maximum oscillation power $\nu_{max}$:
\begin{equation}
\Delta \nu=\sqrt{\frac{M/M_{\odot}}{(R/R_{\odot})^3}}134.9 \mu\mathrm{Hz}
\end{equation}
and
\begin{equation}
\nu_{\mathrm max}=\frac{M/M_{\odot}}{(R/R_{\odot})^2\sqrt{T_{eff}/5777 K}}3.05 mHz.
\end{equation}
C-D diagram and scaling laws are usually adopted
to carry out what might be called 'ensemble Asteroseismology' (e.g., Chaplin et al. 2011), a study of
similarities and differences in groups of few hundreds to thousands of stars, such as field stars or clusters.
These tools allow to determine fundamental properties of the studied stars, particularly mass and radius with 5-7 \% of
uncertainty, and age with an
error up to 20\%, depending on the precision with which the small separation and the metallicity are known.

More accurate and precise determination of fundamental parameters of stars, with an uncertainty of 2\% in radius and mass and less than 5\% for age, can
be obtained only using set of individual pulsation frequencies through
computation of stellar models which reproduce the observed seismic properties
of a star (see, e.g., Mathur et al. 2012, Metcalfe et al 2012). 

One of the most important results has been the introduction of the model's refinements obtained by Helioseismology into the stellar modelling in order to lead the model to be consistent with observed pulsations, such as the improvements in the equation of state and the treatment of the elemental diffusion during the evolution. The consequence has been to have greatly improved the agreement between theoretical and observed 
globular cluster magnitude-colour diagrams and removed the historical gap between the age of the universe deduced from cosmology and stellar evolution.

\begin{figure}[b]
\hspace{-0.5cm}
\includegraphics[width=1.1\linewidth]{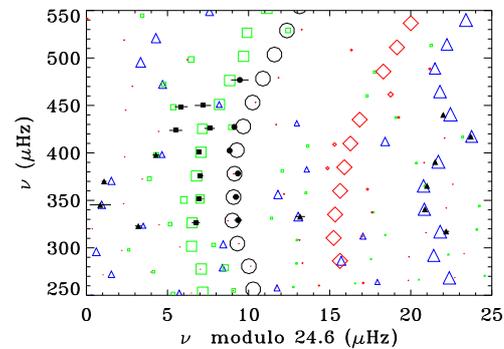}
\caption{\footnotesize \`Echelle diagram based on observed (filled 
symbols) and computed frequencies (open symbols) for the red-giant star KIC4351319 (Di Mauro et al. 2011). 
Circles are used for modes with $l=0$, triangles for $l=1$, squares for $l=2$, diamonds for $l=3$.      
The size of the open symbols indicates the relative surface amplitude of oscillation of the modes.}
\label{eche}   
\end{figure}

A detailed comparison between the theoretical oscillation spectra
and the theoretical oscillation frequencies can be obtained by the \'echelle diagram \citep[e.g.,][]{grec83}
based on Eq. 1 and the asymptotic properties of the oscillation spectrum (see Fig. \ref{eche} obtained for a red giant).
This diagram shows 
for each harmonic degree vertical ridges of frequencies equally spaced by the large separations. The frequency's distance between two adjacent columns of frequencies represents the small separation. 
Oscillation frequencies which do not follow the asymptotic relation, as in the case of gravity or mixed modes, show  a significant departure from the regular pattern (see modes for $l=1$ in Fig. \ref{eche}).

In addition 
the \'echelle diagram shows, for each $l$, that a weak oscillatory motion is present in the observed and calculated frequencies for the acoustic modes which follow closely the asymptotic law.
This important property of the oscillation spectra rises
from sharp variations of the internal stratification occurring at certain acoustic depth in the structure of a star. This signature can be isolated and used, for example, 
to find the location of the base of the convective envelope
and of the region of the second helium ionization (e.g., Miglio et al. 2010; Mazumdar et al. 2012).

\subsection{Seismology of red-giant stars}

The presence of solar-like oscillations in red giants, firstly discovered by the space mission MOST \citep{barban07}, was well established by the CoRoT satellite
\citep{deridder09}, which was able to find
solar-like oscillations in a very large sample of G and K giant stars \citep[e.g.,][]{hekker09} mainly lying in the core-helium-burning evolutionary phase.  

The high-quality observations of the {\it Kepler} mission  enabled to detect solar-like oscillations in more than 1000 
red-giant stars from the red clump to the lower luminosity region of the
 red-giant branch \citep[e.g.,][]{bedding10}, where stars are still burning H in the shell.
 
The properties of solar-like oscillations are expected to change
as the stellar structure evolves. 
According to Eq.~(\ref{eq1}) and considering that $\Delta\nu \propto R^{-3/2}$,
oscillation frequencies of a given harmonic degree should decrease as the star evolves and 
the radius increases   
and should be almost uniformly spaced by $\Delta \nu$ at
each stage of evolution. However, in subgiants and red giants the   
radial modes seem to follow Eq.~(\ref{eq1}) closely, but
the frequencies of some non-radial modes appear to be shifted from the 
regular spacing due to the occurrence of the so-called `avoided crossing' \citep{dim03}.
As the star evolves away from the main sequence, the core contracts 
and the radius expands, causing an increase of the local gravitational 
acceleration and of the gradients in the hydrogen abundance, and hence of
the buoyancy frequency in the deep interior of the star. 
As a consequence g modes with high frequencies
are allowed to propagate and can interact with p modes of similar frequency
and same harmonic degree, giving rise to modes with mixed character, 
which behave as g modes in the interior and p modes in the outer envelope 
\citep{aizenman77}.
The interaction can be explained as the coupling of two oscillators of similar frequencies.
The effect of 
the coupling becomes much weaker for modes with higher harmonic degree,
since in these cases the gravity waves are better trapped in the 
stellar interior and hence better separated from the region of propagation 
of the acoustic waves \citep{dziembowski01}. 

It has been found by
\citet{montalban10} and observationally demonstrated by \citet{huber10}, that the scatter 
of $l=1$ modes caused
by `avoided crossing' decreases  as the star goes up to the red-giant branch:
as the luminosity increases and the core become denser, the $l=1$ acoustic
modes are better trapped and the oscillation spectra become more regular.
Once the star ignites He in the core, the core expands and the external 
convective zone becomes shallower which has the effect of increasing the probability of
coupling between g and p modes again.

 Very recently, \citet{beck11} have demonstrated that the quality of the {\it Kepler} observations gives the possibility to
measure the period spacings of mixed-modes with gravity-dominated character 
which, like pure gravity modes, penetrate deeply in the core allowing
 to study the density contrast between the core region and the convective envelope and, like p modes, have amplitude at the surface high enough to be observed.
In particular, 
\citet{bedding11} found that measurements of the period spacings of the
 gravity-dominated mixed modes,
permit to distinguish between different stages of evolution:
the hydrogen-burning stage having a $\Delta P\simeq 50$s and the helium-burning phase with a $\Delta P\simeq 200$s.

The occurrence of mixed modes is then a strong indicator of the evolutionary state of a star and the fitting of the observed modes with those calculated by theoretical models can provide not only mass and radius but, with a good approximation, an estimate of the age of a red-giant star (e.g., Di Mauro et al. 2011).
\subsection{Dynamics of red-giant stars}

Accurate observations of rotational splittings 
provide, as explained in Sec. 3.1 for the Sun,
 stringent constraints on the internal dynamics of the observed star.

The {\it Kepler} satellite has been able to detect
frequency splittings in several main sequence stars and a large number of red-giants stars. 
In particular, Beck et al. (2012) found 
that in the observed red giants, the splittings of the g-dominated
dipole modes are on average 1.5 times larger than
the splittings of the p-dominated modes. By comparison with theoretical models, the larger g-dominated mode splittings have been interpreted as the signature of a
core rotating at least 10
times faster than the surface. Non-rigid rotation
has also been derived by Deheuvels et al. (2012) in a
young giant at the beginning of the RGB phase by applying seismic inversion techniques to the observed rotational splittings.

\begin{figure}
  \includegraphics[width=1.0\linewidth]{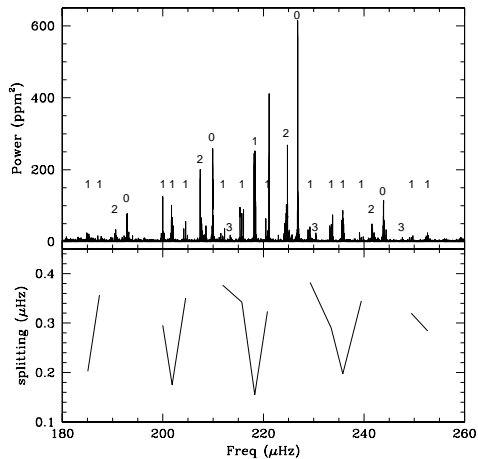}
\caption{\footnotesize The upper panel shows the oscillation spectrum of KIC4448777 observed with {\it Kepler} from Di Mauro et al. (2012). 
The harmonic degree of the observed modes
($l$=0,1,2,3) are indicated. Multiplets due to rotation are
visible for $l$=1. The lower panel shows the values of the
observed rotational splitting for individual $l$=1 modes.}
\label{fig:1}      
\end{figure}

Figure \ref{fig:1} shows
 the oscillation spectrum and the rotational splittings
for a red giant star studied recently by Di Mauro et al. (2012).

\section{Conclusions}


The recent stellar asteroseismic
results, driven by new satellites observations of unprecedented quality and scope, 
have put Helioseismology and the Sun into a broader context, confirming that 
techniques and tools developed for Helioseismology
can be applied with success to other stars.

The golden era of Asteroseismology has indeed just open its windows showing the potential of stellar pulsations for studying
fundamental parameters, such as mass, radius and age of main sequence and also more evolved stars. In particular, the novel striking finding is represented by the discovery 
that red-giant stars can be probed into their deeper interior, as contrary to the main-sequence stars and the Sun itself, thanks to the use of the mixed modes which can be measured at surface. 
Furthermore, having proved to be able to measure 
the core's rotation in evolved stars, 
it appears not far the moment in which it will be possible to understand
how stellar evolution modifies rotational properties and
the angular momentum is conserved 
 as
a star advances towards the helium-core burning phase.

Asteroseismology provides, without doubts, an extensive possibility for testing and understanding the physical processes occurring in a star and there is high expectation from the 
data that will be obtained in the next future for stars 
with different pulsational characteristics.
It is evident that the resulting improvements in stellar characterization and modelling will be indeed crucial for broad areas of astrophysics, including the investigation of the structure and evolution of the Galaxy and the understanding of the formation of elements in the Universe.

\bibliographystyle{aa}

\end{document}